# Magnon-assisted tunnelling in van der Waals heterostructures based on CrBr$_3$

D. Ghazaryan[1], M.T. Greenaway[2], Z. Wang[1], V.H. Guarochico-Moreira[1,3], I.J. Vera-Marun[1], J. Yin[1], Y. Liao[1], S. V. Morozov[4], O. Kristanovski[5], A. I. Lichtenstein[5], M. I. Katsnelson[6], F. Withers[7], A. Mishchenko[1,8], L. Eaves[1,9], A. K. Geim[1,8], K. S. Novoselov[1,8,*], A. Misra[1]

[1]*School of Physics and Astronomy, University of Manchester, Oxford Road, Manchester, M13 9PL, UK*

[2]*Department of Physics, Loughborough University, Loughborough, LE11 3TU, UK*

[3]*Escuela Superior Politécnica del Litoral, ESPOL, Facultad de CNM, Campus Gustavo Galindo Km 30.5 Vía Perimetral, P.O. Box 09-01-5863, Guayaquil, Ecuador*

[4]*Institute of Microelectronics Technology and High Purity Materials, RAS, Chernogolovka, 142432, Russia*

[5]*Institute of Theoretical Physics, University Hamburg, D-20355, Hamburg, Germany*

[6]*Institute for Molecules and Materials, Radboud University, 6525AJ, Nijmegen, Netherlands*

[7]*University of Exeter, College of Engineering, Mathematics and Physical Sciences, Exeter EX4 4SB, Devon, England*

[8]*National Graphene Institute, University of Manchester, Oxford Road, Manchester, M13 9PL, UK*

[9]*School of Physics and Astronomy, University of Nottingham, Nottingham, NG7 2RD, UK*

**The growing family of two-dimensional (2D) materials[1-3] can be used to assemble van der Waals heterostructures with a wide range of properties[4-6]. Of particular interest are tunnelling heterostructures[7-9], which have been used to study the electronic states both in the tunnelling barrier and in the emitter and collector contacts[10,11]. Recently, 2D ferromagnets have been studied theoretically[12-15] and experimentally[16-18]. Here we investigate electron tunnelling through a thin (2-6 layers) ferromagnetic CrBr$_3$ barrier. For devices with non-magnetic barriers, conservation of momentum can be relaxed by phonon-assisted tunnelling[8,19-21] or by tunnelling through localised states[8,21,22]. In the case of our ferromagnetic barrier the dominant tunnelling mechanisms are the emission of magnons[18] at low temperatures or scattering of electrons on localised magnetic excitations above the Curie temperature. Magnetoresistance in the graphene electrodes further suggests induced spin-orbit coupling and proximity exchange via the ferromagnetic barrier. Tunnelling with magnon emission offers the possibility of spin-injection, as has been previously demonstrated with other ferromagnetic barriers[23,24].**

Our devices were assembled on an oxidised silicon wafer by the dry transfer method in an inert atmosphere. The layer structure of our devices is Si/SiO$_2$/hBN/Gr/CrBr$_3$/Gr/hBN; here Gr stands for graphene, see Fig. 1a. The typical area of the devices is a few tens of square microns. We prepared a number of devices with different numbers N=2, 4 and 6 of CrBr$_3$ monolayers. A tunnel current, $I$, flows when a bias voltage, $V_b$, is applied across the two graphene layers. A typical plot of differential conductance $G=dI/dV_b$ is presented in Fig. 1b. The zero-bias conductance depends exponentially on the barrier thickness, see Fig. 1c.

The dependence of $G$ on the applied gate ($V_g$) and bias voltages for a sample with 6 CrBr$_3$ monolayers is presented as a colour map in Fig. 1b. In previously studied devices with hBN barriers, the passage of the Fermi level through the zero in the density of states at the Dirac points of the top and bottom graphene electrodes is observed as an X-shaped feature in the colour map[10,11,19,25], where $G=0$. This feature is not observed in the devices with a CrBr$_3$ barrier. We attribute this to a high level of doping of the graphene electrodes by the CrBr$_3$.

A magnetic field, $B_\perp$ applied perpendicular to the layers (i.e. parallel to the tunnel current) causes the electron spectrum to quantise into a series of Landau levels (LL). In Fig. 1d the Landau quantisation is revealed in $G(V_b, V_g)$ as a series of parallel stripes, with positive slope, due to the passage of the chemical potential through the gaps between LLs. Only this series of peaks with a positive slope is observed, corresponding to Landau quantization in the lower graphene electrode. The absence of stripes with a negative slope that would arise from LLs in the top electrode indicates that the CrBr$_3$ layer screens the gate voltage-induced electric field, so that the chemical potential in the top graphene layer is unchanged by $V_g$.

Figures 1b,e reveal a series of step-like increases of the tunnel conductance, with values of $V_b$ that are independent of the gate voltage. This type of behaviour is characteristic of inelastic tunnelling with phonon emission and has been observed previously for Gr/hBN/Gr tunnel junctions[8,19-21]. When the two graphene electrodes of the device are crystallographically misaligned, in-plane momentum conservation requires that the tunnelling electrons are scattered by impurities or by the emission of a phonon[10].

In order to identify the quasiparticles responsible for inelastic tunnelling events in the CrBr$_3$ devices, we measure $G$ as a function of magnetic field parallel to the layers, $B_\parallel$, (perpendicular to the tunnelling current), see Fig. 1e. In this geometry LL quantisation of the electrons in graphene is absent. The measured resistance of the device contacts is practically independent of applied magnetic field; therefore, any change in position of a particular feature in conductance must be due to a change in the tunnel mechanism. In Fig. 1e,f some of the peak positions are clearly dispersed in $V_b$ by the applied magnetic field, whereas others are not. We attribute the non-dispersed peaks to phonon-assisted tunnelling[8,19-21] and the dispersed peaks to magnon-assisted tunnelling[18,26]. The dispersing peaks are shifted linearly with $B_\parallel$, with a slope of between $(5.1\pm0.2)\cdot\mu_B$ and $(5.7\pm0.2)\cdot\mu_B$, depending on the particular peak, here $\mu_B = 5.79\times10^{-5}$ eV/T is the electron Bohr magneton.

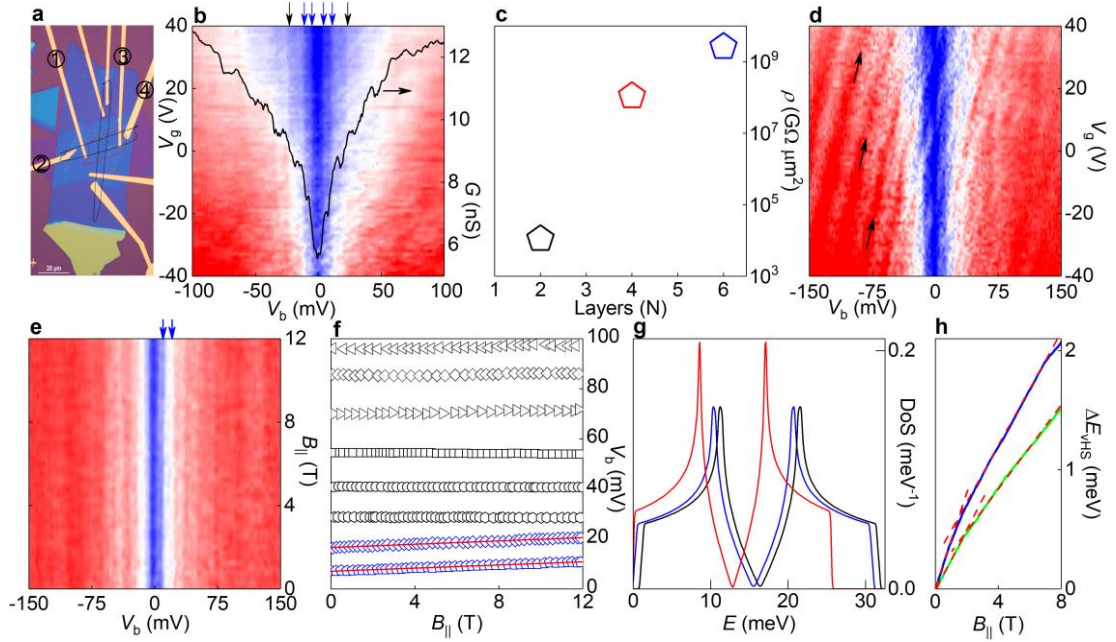

**Figure 1| Inelastic magnon assisted tunneling. a,** Optical micrograph of the investigated device. Black and red contours outline the graphene and CrBr$_3$ flakes, respectively. Scale bar - 20μm. **b,** Zero-field differential tunneling conductance $G$ dependence on the gate and bias voltages (colour scale is blue to white to red, 6nS to 11.5nS to 17nS) for a 6 layer CrBr$_3$ tunnel barrier device. Black curve is the tunneling conductance when $V_g$=0V. The arrows indicate some of the step-like increase in $G$ which corresponds to inelastic tunneling with the emission of a magnon (blue) phonon (black). **c,** Zero bias resistivity of junctions with different barrier thicknesses. **d,** as b, in perpendicular field of $B_\perp$=12.5T. Black arrows correspond to peaks in conductance due to LL spectrum. Colour scale is blue to white to red, 6nS to 12nS to 19nS. **e,** Differential tunneling conductance $G$ as a function of $B_\parallel$ and $V_b$ ($V_g$=0V). Colour scale is blue to white to red, 6nS to 12nS to 19nS. **f,** Bias position of the step-like features in $G$ as a function of $B_\parallel$. Red lines are the linear fitting with the slopes (from bottom to top) (5.7±0.2)·μ$_B$, (5.1±0.2)·μ$_B$. **g,** Calculated magnon density of states for $T$=10K, $B$=0T (blue line), $T$=10K, $B$=6.25T (black line), $T$=$T_C$, $B$=0T (red line). Same calculations provide $T_C$=88K. **h,** Calculated changes of the position of the van Hove singularities in magnon density of states (g) as a function of magnetic field for temperatures close to $T_C$. Green curve – for the peak at 8.5meV at $B$=0T, blue – 17meV at $B$=0T. Red dashed lines are the guides to the eye to highlight the change of slope at low magnetic fields.

The calculated density of magnon states in CrBr$_3$, based on experimentally measured exchange parameters[27,28], is shown in Fig 1g. The two van Hove singularities in the spectrum at 8.5 meV and 17 meV arise from the hexagonal symmetry of the CrBr$_3$ lattice. We relate these peaks to the two step-like increases in $G$ observed in the experiment at low bias when $V_b$≈7.5 mV and 17mV (see Fig. 1b). Previously reported inelastic neutron scattering studies of magnons in CrBr$_3$ are in a good agreement with our calculations and indicate that the magnon energy is limited to about 30meV[27,28]. This supports our assumption that non-dispersing steps in $G$ at energies above 30meV should be attributed to phonon-assisted tunnelling.

In a magnetic field, the Zeeman effect would shift the whole magnon spectrum by an energy 2μ$_B$B, along with the two van Hove singularities, if the magnon-magnon interaction is neglected (Fig. 1g). In the regime of low temperatures and high magnetic fields, self-consistent spin-wave calculations[29] predict a larger shift of the van Hove singularities of ~2.4μ$_B$B. This enhanced shift of the magnon spectrum is due to magnon-magnon interactions, as explained in detail in the Supplementary Information. However, the observed magnetic field-induced shift of the position of the steps in conductance of ~5μ$_B$B, is approximately twice larger than that expected by our theory. Currently, we have no explanation for this discrepancy; understanding it will be the subject of future work. However, we note that at low $B$ and temperatures close to $T_C$, our self-consistent spin-wave calculations, predict a magnetic field-induced shift in the two van Hove singularities of 4.5μ$_B$B and 7.1μ$_B$B respectively (Fig. 1h and Supplementary Information) which is comparable to the shift measured in our experiments. Such a strong dependence of magnon spectra on magnetic field close to $T_C$ can be understood as follows. Magnon frequencies depend strongly on temperature because both long- and short-range magnetic order are suppressed by spin fluctuations: both interlayer exchange interactions and the Zeeman gap in the magnon spectrum suppress the fluctuations, making the magnons stiffer. The change of the regime corresponds to the condition 2μ$_B$B≈$J_L$ ($J_L$ is the exchange integral between the layers) which is indeed seen as the change of the slope on the calculated dependence of the energy position of the van Hove singularities on the magnetic field at low $B$, Fig. 1h.

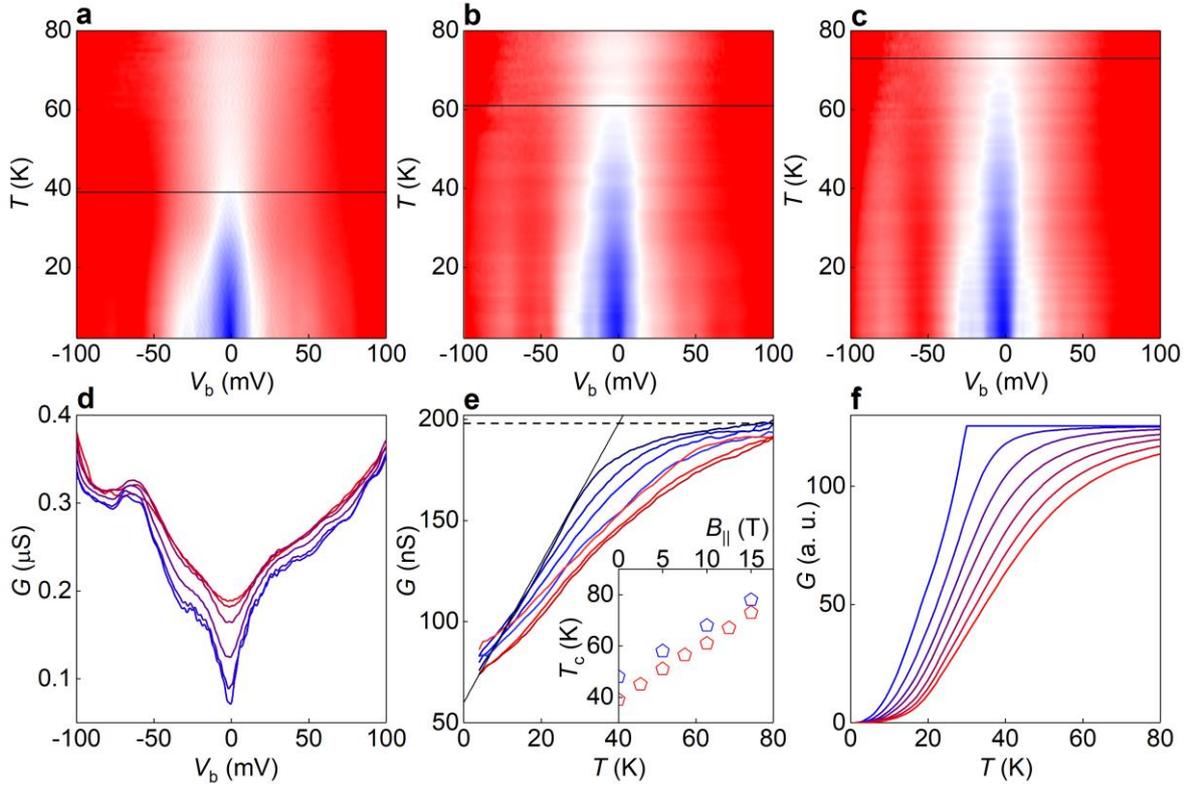

**Figure 2| Effect of conductance on in-plane magnetic field. a-c,** $G(V_b,T)$ when $V_g$=0V and $B_\parallel$=0T (a), $B_\parallel$=10T (b), and $B_\parallel$=15T (c) for a device with 4 layer thick CrBr$_3$ tunnel barrier (colour scale blue to white to red, 0.06μS to 0.17μS to 0.3μS for all contour plots in the figure). Black horizontal lines in a,b,c mark the extracted value of $T_C$. **d,** Zero-field tunneling conductance plots for various temperatures at a fixed gate voltage of $V_g$=0V (blue to red curves range from T=2K to T=50K with the step 10K). **e,** Dependence of the zero bias tunneling conductance $G$ on $T$ for different $B_\parallel$ (blue to red curves range between $B_\parallel$=0T and $B_\parallel$=15T with a 2.5T increment). Straight lines are guide for an eye and illustrate the method for extraction of the Currie temperature (plotted for $B_\parallel$=0T). The dashed black horizontal line illustrates the saturating conductance at the high temperatures, and the straight solid line indicates the tangent to the particular curve. Inset: The extracted dependence of the $T_C$ on $B_\parallel$. Red and blue symbols are for devices with the thickness of the CrBr$_3$ barrier 4 and 6 layers respectively. **f,** Theoretical modelling of the zero bias tunneling conductance $G$ dependence on $T$ for different $B_\parallel$ (blue to red curves ranges between $B_\parallel$=0T and $B_\parallel$=15T with a 2.5T increment).

We now consider the dependence of the differential conductance on temperature, $T$. Fig. 2a,d plot $G$ as a function of $V_b$ and $T$. We observe that the zero-bias differential conductance, $G_0$=$G(V_b$=0), increases with increasing temperature. Its rate of change, d$G_0$/d$T$, is largest for temperatures between 15-30K and saturates above 40K, close to the Curie temperature, $T_C$≈35K[27,30], for CrBr$_3$, Fig. 2e. This behaviour is similar for devices with different CrBr$_3$ barrier thicknesses (see Supplementary

Information) and we attribute it to an increase of elastic scattering of tunnelling electrons with temperature. We also measure the dependence of $G$ on bias voltage and temperature for different $B_∥$, Fig. 2a-c. In all cases $dG_0/dT$ decreases with increasing $B_∥$, Fig. 2e.

At temperatures close to $T_C$ the magnon population is very high, so absorption of a magnon by a tunnelling electron is more probable. Also, at such temperatures the short-range magnetic ordering *between* the layers and the long-range magnetic ordering *within* the CrBr$_3$ layer are destroyed as described by the self-consistent spin wave theory of quasi-two-dimensional magnets[29,31]. The correlation length $\xi(T)$ within the layers is estimated to be

$$\frac{a}{\xi(T)} = \sqrt{\frac{2|\mu|}{3J\gamma}}$$

where $a$ is the nearest-neighbour distance between magnetic atoms, and $\mu$ is the chemical potential of the self-consistent spin waves (see Supplementary Information). $J\gamma$ is the renormalized in-plane exchange interaction and the renormalisation parameter γ is determined from self-consistent equations and defines the magnon hopping between adjacent sites, see Supplementary Information and reference [29]. It also can be defined as γ=$SJ_{eff}/J$, where $S$=3/2 is the spin of Cr ion and $J_{eff}$ is the effective exchange parameter in finite temperatures and magnetic fields when magnon occupation factors are finite.

These observations suggest two possible mechanisms for elastic tunnelling with momentum conservation at zero bias. First, consider a two-magnon process. In the case of a ferromagnetic barrier, emission of two magnons by a tunnelling electron is forbidden by spin conservation. However, a three-particle process, in which an electron emits one magnon and absorbs another, is possible. This type of process does not change the energy of the electron, changing only its momentum, thus ensuring momentum conservation for electrons tunnelling between the Dirac cones of the two misoriented graphene layers. The ratio between the intensities of the magnon emission (Stokes) and absorption (anti-Stokes) processes is given by exp($E_m/k_BT$), where $E_m$ is the magnon energy and $k_B$ is the Boltzmann constant. At low temperatures, absorption processes are suppressed. However, close to $T_C$ and for our range of parameters, the typical magnon energies is of $\sim k_BT_C$. In 3D magnets with $S$≈1, typical magnon energies are of the order of $k_BT_C$ since they are both proportional to $zJ$, where $z$=3 is the nearest-neighbor number. For 2D magnets, $T_C$ is suppressed by a factor $\sim 1/\ln(J/J_L)$ (see[29] for explicit relations). The exchange integrals for CrBr$_3$ have been previously extracted experimentally[27,28] as $J$=1.698mV and $J_L$=0.082mV, giving the $J/J_L$≈21 and the suppression factor to be ~1/3, still of the order of unity. This makes the three particle Stokes-anti-Stokes processes significant.

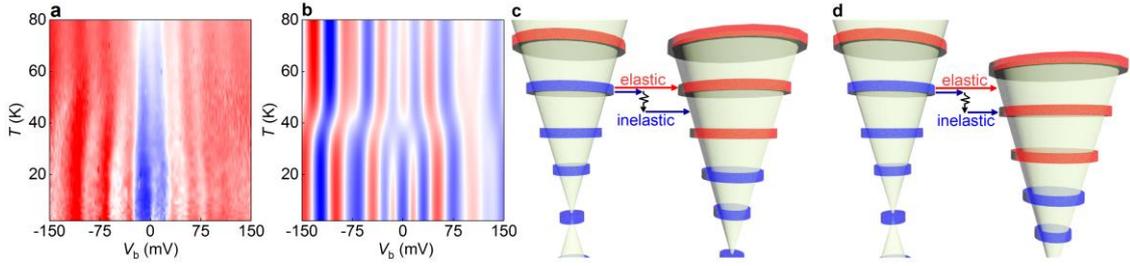

**Figure 3| Inter-LL tunnelling. a,** measured and **b,** calculated $G(V_b,T)$ when $V_g=0V$ and $B_\perp=17.5T$ for a device with 6 layer $CrBr_3$ barrier (colour scale blue to white to red, 4nS to 13nS to 22nS). **c,d,** Schematic diagrams of inelastic and elastic tunnelling events between the LL spectrum of the two graphene layers. Blue states are occupied, red – empty. At low temperatures the tunnelling is inelastic (blue arrows) with magnon emission (black arrow). Increasing temperature increases the contribution to the current of elastic two magnon and spin disorder scattering (red arrows). Application of a bias voltage differentiates between these two processes, in c the bias is such that elastic processes allows resonant tunnelling between the 3$^{rd}$ and 4$^{th}$ LLs. In d the bias is higher, which brings the same 3$^{rd}$ and 4$^{th}$ LLs into resonance, but is now mediated by inelastic tunnelling.

A second potential elastic scattering mechanism, which becomes dominant for $T\sim T_C$, is electron scattering on the imperfections in the spin texture within a $CrBr_3$ layer[32,33]. These imperfections break translational symmetry and are capable of scattering the momentum of electron by $1/\xi(T)$. According to our calculations, for temperatures close to $T_C$, the correlation length falls to values $\sim 10a$, and to $\sim 2a$ at $T\sim 1.5 T_C$ (see the Supplementary information). This length scale is much smaller than the de Broglie wavelength of the tunnelling electrons, which is of the order of tens or even hundreds interatomic distances for the case of graphene under the measurement conditions reported here. Hence close to and above $T_C$ tunnelling electrons encounter a highly disordered spin configuration; short range order is lost entirely for $T>T_C$[31]. Scattering of electrons on the short-range spin disorder is therefore analogous to the electron scattering on charged impurities, thereby ensuring momentum conservation for electrons tunnelling between misaligned graphene layers.

We model the elastic tunnelling rate at zero bias by calculating the scattering rate due to two magnon processes at low temperatures ($T \lesssim T_C$) and scattering arising from spin imperfections in the ferromagnet at high temperatures ($T \gtrsim T_C$). For $T \lesssim T_C$, elastic scattering is dominated by two-magnon processes with a rate, $W_{2mag}$, proportional to the contribution of the longitudinal spectral density[34,35] (also see Supplementary Information). The probability of these two-magnon processes would go to zero above $\sim T_C$ because the magnon description of spin excitation become invalid. Instead, for $T \gtrsim T_C$, the elastic tunnelling rate, $W_{spin}$, is dominated by scattering on the disordered spin texture[32,33], and is therefore dependent on the magnetisation of the ferromagnet[36]. We combine the two scattering rates by smoothly varying the amplitudes of the two scattering rates to reflect the increase in the thermal spin disorder of the lattice, so that $W_{tot} = (1- f(T))\ W_{2mag} + f(T)\ W_{spin}$, where $f(T) = 0.5[1+\tanh(\beta(T/T_C-1))]$. We obtain a good fit to the data when $\beta=4$, see Fig. 2f.

The measured magnetic field-dependence of the conductance also fits well to our model: as the external magnetic field increases, the magnon energy increases due to the Zeeman effect (Fig. 1g), thus reducing the magnon population and the rate $W_{2mag}$ at a given $T$. At high temperatures, the external magnetic field contributes to spin alignment, thus reducing scattering by imperfections of the spin texture.

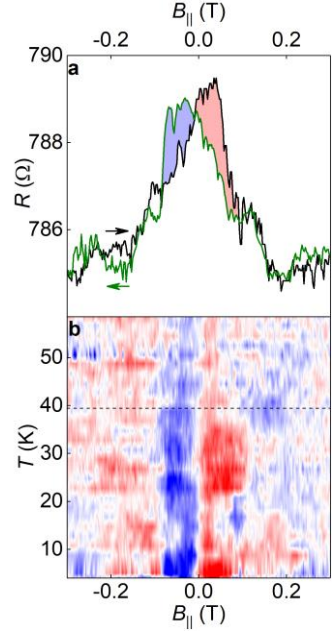

**Figure 4| Magnetotransport in graphene proximitised with CrBr$_3$. a,** Four probe resistance plots of the bottom graphene electrode in a device with 4 layer thick CrBr$_3$ barrier as a function of in-plane magnetic field $B_\parallel$, at $T$=5 K. The black and the green curves correspond to sweeping $B_\parallel$ in the trace and retrace directions, respectively, indicated by the corresponding arrows. In the low field regime, $|B_\parallel|$ < 0.1 T, the graphene exhibits a magnetoresistance of 0.5% and a hysteretic behaviour. The latter is indicated by the blue and red coloured areas between the trace and retrace curves. **b,** Colour map of the difference between trace and retrace curves as in a, now also as a function of temperature (colour scale blue to white to red, -1 Ω to 0 Ω to 1 Ω). Black horizontal dash line marks the extracted value of $T_C$, as in Fig. 2a. The hysteresis is visible in the colour map as the blue and red lobes in the range $|B_\parallel|$ < 0.1 T. Both the magnitude of the hysteresis and the range of the hysteresis in $B_\parallel$ decrease for $T>T_C$, indicating their origin from the long-range magnetic order in the CrBr$_3$ barrier.

We also consider the effect of $B_\perp$ on $G$ as a function of bias and temperature. Tunnelling between the LLs of the two graphene layers gives rise to peaks in $G$, see Fig. 3a. As $T$ approaches $T_C$ we observe a clear shift in the position of the peaks in $V_b$. We interpret this behaviour as a transition from inter-LL tunnelling dominated by inelastic transitions to tunnelling dominated by elastic transitions, as illustrated in the band diagrams, Fig. 3c and d, where inelastic and elastic tunnelling processes are depicted by blue and red arrows respectively.

For instance, for a process depicted on Fig. 3c, in case of inelastic tunnelling, an electron in the 3$^{rd}$ LL of the bottom electrode cannot tunnel with the emission of a magnon since its reduced energy lies in the energy gap between the 3$^{rd}$ and 4$^{th}$ LLs of the top graphene electrode, see blue arrow in Fig. 3c. In contrast, a peak in conductance is observed for electrons tunnelling with conservation of energy, red arrow in Fig. 3c, because the 3$^{rd}$ LL in the bottom electrode is aligned in energy with the 4$^{th}$ LL in the top electrode. A slightly higher bias voltage brings into alignment the reduced energy of the inelastic tunnelling electron with an unoccupied state in the 4$^{th}$ LL in the top electrode, resulting in a peak in conductance, Fig 3d. In order to demonstrate the effect of the temperature-dependent change in the relative contributions of these two processes, we show in Fig. 3b a colour map of conductance calculated using a model developed previously for phonon-assisted[19] inter-LL tunnelling[25] in a device with a non-magnetic hBN barrier. The model includes the two competing processes: inelastic tunnelling processes induced by a single magnon with an energy of 10 meV and a scattering amplitude, $\propto 1- f(T)$, which decreases as the temperature increases through $T_C$ ; and elastic tunnelling processes whose amplitude, $\propto f(T)$, increases as temperature increases. The model is in qualitative agreement with measured data, see Fig. 3a,b.

This type of ferromagnetic tunnel barrier could be used to inject spin polarized carriers into graphene or other 2D materials. Pioneering work in traditional 3D materials has established the possibility to create a large (close to 100%) spin polarization in the tunnel current using ferromagnetic barriers, quantified via tunnelling spectroscopy using a superconductor electrode (also known as the Meservey-Tedrow technique)[23,24]. In van der Waals heterostructures based on graphene and other 2D crystals the mechanisms of spin injection, and thus the possibility of obtaining large spin polarisation, are linked strongly to the momentum conservation conditions. Thus, in the case when the tunnelling states in the emitter and collector are displaced in momentum space (as for twisted graphene lattices or for two materials with different lattice periodicity), conservation of in-plane momentum can be enabled by magnon emission, which selects only one spin polarization. Note that this is not the case for two magnons processes, which cannot provide spin polarisation. The use of ferromagnetic tunnel barriers could therefore enrich the type and functionality of van der Waals heterostructures.

A quite different set of physical phenomena, with the potential to enable van der Waals heterostructures for spintronics, are proximity effects between electrode and ferromagnetic barrier. Recent work have provided evidence for the presence of both an induced exchange field and spin-orbit coupling in graphene, when in contact with a magnetically ordered 3D insulator, either by measuring the anomalous Hall effect[37] or by directly detecting pure spin current[38]. There is also evidence that proximity effects can lead to anisotropic magnetoresistance in graphene[39]. In our multi-terminal tunnelling structures, we have also measured in-plane transport *within* each of the two graphene electrodes. Thus, for the device shown in Fig. 1a we measure the voltage between contacts 1 and 3, when a current flows from contacts 2 to 4 in the graphene layer below the ferromagnetic barrier (similar measurements were made for the top electrode with current flowing above the ferromagnetic barrier). In this type of measurement the transport within a single graphene electrode is studied as a function of the in-plane magnetic field, for which there is neither Lorentz magnetoresistance nor LLs quantization. Surprisingly, we still observe an appreciable magnetoresistance of ~0.5%, Fig. 4a. More importantly, the magnetoresistance exhibits hysteresis on the in-plane magnetic field, with this hysteresis diminishing rapidly for temperatures above $T_C$, Fig. 4b.

We attribute these observations to a proximity-induced anisotropic magnetoresistance (PAMR) in graphene[40], driven by the magnetization reversal processes of the tunnel barrier. The PAMR response is in qualitative and quantitative agreement with recent experimental[39] and theoretical[40] works, and indicates the presence of both spin-orbit coupling and an exchange field[41] in the graphene electrode. Finally, we note that the observation of PAMR provides an independent demonstration of ferromagnetic order in the tunnel barrier. This is further evidence that the vertical transport effects shown in Fig. 2 originate from the magnetic nature of the barrier.

We argue that tunnelling with magnon emission and proximity effects offers new prospects for exploiting 2D ferromagnetic barriers in graphene spintronics. In particular, the tunnel current can be spin-filtered. For the case of tunnelling through a ferromagnetic barrier, one-magnon processes will result in selective tunnelling of only one spin polarisation. Indeed, at low temperatures spin conservation, combined with the emission of a magnon, enables a tunnelling electron, initially in the spin-down state, to flip its spin into a final spin-up state. The one-magnon tunnelling process from an initial spin-up state is forbidden. Two-magnon processes would allow both spin polarisations to tunnel. However, they have a small probability at low temperatures since they require not only emission but also absorption of magnons; there are no magnons available for absorption near the ferromagnetic ground state.

**Comments**

During the preparation of the manuscript the authors became aware of similar work done using a different chromium-trihalide (CrI3) [42-44].

**Methods**

**Fabrication**

The $CrBr_3$ crystals were purchased from a commercial supplier, HQ graphene.

$CrBr_3$ was first exfoliated onto a polypropylene carbonate (PPC) coated $SiO_2$/Si substrate in an argon-filled glove box maintaining water and oxygen concentration less than 0.1ppm. $CrBr_3$ flakes of different thicknesses (2 to 6 layers) were identified by using the contrast variation under different colour filters and dark field imaging. Heterostructures comprising SLG/$CrBr_3$/SLG encapsulated within thin boron nitride (BN) layers were assembled on a 290 nm $SiO_2$/Si substrate following standard dry transfer procedure using a PMMA membrane. First, a single layer graphene (SLG) was picked up by a thin boron nitride layer (~8-10 nm BN) using the PMMA membrane. Subsequently, a suitable $CrBr_3$ layer was picked up by SLG/BN/PMMA membrane in the glove box. Finally, this stack of $CrBr_3$/SLG/BN on the PMMA membrane was peeled onto the SLG/BN/290 nm $SiO_2$/Si substrate. The final stack consists of Si/ 290nm $SiO_2$/ ~25nm BN/ SLG/ CrBr3/ SLG/ ~ 10nm BN.

For electrical characterization of the tunnelling devices, Cr/Au edge contacts were made on the top and bottom graphene layers using electron beam lithography followed by boron nitride etching, metal deposition and a lift-off process. Boron nitride was etched in a reactive ion etching system using $CHF_3$ and oxygen chemistry. Contacts on the graphene sheets were made so as to have a four probe measurement geometry. Note that during the heterostructure assembly, the top and bottom graphene layers were chosen to extend beyond the $CrBr_3$ layer so that the contact processing (especially top BN etching and subsequent lift-off process) does not affect the $CrBr_3$.

**Modelling**

The magnon density of states was calculated in the nearest-neighbour approximation with the values of exchange integrals $J$=1.698meV and $J_L$=0.082meV taken from the experiment[27,28]. The temperature and magnetic field dependences of the magnon spectra were calculated within the self-consistent spin-wave theory[27]. The details are presented in the Supplementary Information.

**Acknowledgements**

This work was supported by the EU Graphene Flagship Program, the European Research Council Synergy Grant Hetero2D, the Royal Society, the Engineering and Physical Research Council (UK), the US Army Research Office (W911NF-16-1-0279). S.V.M. was supported by RFBR (17-02-01129a) and RAS Presidium Program N4 (task 007-00220-18-00).

# Supplementary Information
# Magnon-assisted tunnelling in van der Waals heterostructures based on CrBr$_3$

*D. Ghazaryan[1], M.T. Greenaway[2], Z. Wang[1], V.H. Guarochico-Moreira[1,3], I.J. Vera-Marun[1], J. Yin[1], Y. Liao[1], S. V. Morozov[4], O. Kristanovski[5], A. I. Lichtenstein[5], M. I. Katsnelson[6], F. Withers[7], A. Mishchenko[1,8], L. Eaves[1,9], A. K. Geim[1,8], K. S. Novoselov[1,8,\*], A. Misra[1]*

[1]School of Physics and Astronomy, University of Manchester, Oxford Road, Manchester, M13 9PL, UK

[2]Department of Physics, Loughborough University, Loughborough, LE11 3TU, UK

[3]Escuela Superior Politécnica del Litoral, ESPOL, Facultad de CNM, Campus Gustavo Galindo Km 30.5 Vía Perimetral, P.O. Box 09-01-5863, Guayaquil, Ecuador

[4]Institute of Microelectronics Technology and High Purity Materials, RAS, Chernogolovka, 142432, Russia

[5]Institute of Theoretical Physics, University Hamburg, D-20355, Hamburg, Germany

[6]Institute for Molecules and Materials, Radboud University, 6525AJ, Nijmegen, Netherlands

[7]University of Exeter, College of Engineering, Mathematics and Physical Sciences, Exeter EX4 4SB, Devon, England

[8]National Graphene Institute, University of Manchester, Oxford Road, Manchester, M13 9PL, UK

[9]School of Physics and Astronomy, University of Nottingham, Nottingham, NG7 2RD, UK

**S1. Device fabrication**

**S2. Temperature dependence of differential *dI/dV*$_b$ conductance on magnetic field for devices with different thickness of *CrBr*$_3$**

**S3. Quantum capacitance of *Gr/CrBr*$_3$*/Gr* devices**

**S4. Calculation of magnon density of states**

**S5. Scattering rates**

**S1. Device fabrication**

Several devices with a different number of CrBr$_3$ layers (2 to 6) were fabricated. Fig. S1 depicts the procedure for fabricating a device with 4 layers of CrBr$_3$.

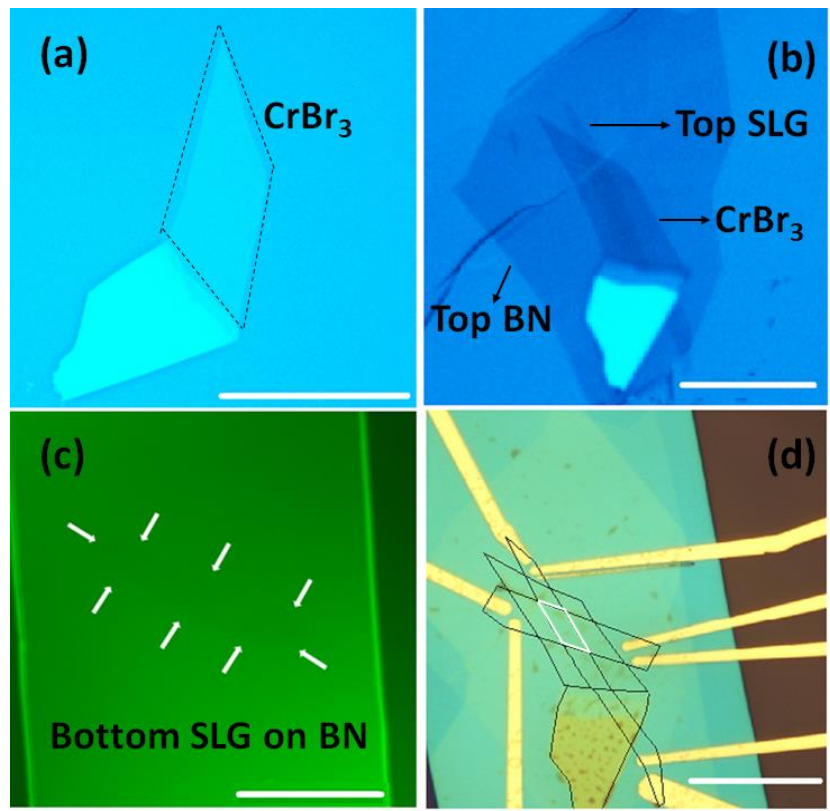

Fig.S1. Optical images of SLG/CrBr$_3$/SLG tunnel device encapsulated with hBN during different fabrication steps. **a,** 4 layers CrBr$_3$ flake (marked by black dashed line). **b,** CrBr$_3$/SLG/top BN picked up on PMMA membrane (SLG – single layer graphene). **c,** SLG (marked by the white arrows) transferred on bottom hBN. After this, the stack shown in b, was transferred on the SLG/BN/SiO$_2$/Si shown in c,. **d,** BN encapsulated SLG/CrBr$_3$/SLG stack with deposited Cr/Au contacts. Black lines represent different layers and white parallelogram marks the actual tunnelling area. Scale bars in all images are 20 µm.

## S2. Temperature dependence of differential $dI/dV_b$ conductance on magnetic field for devices with different thickness of CrBr$_3$

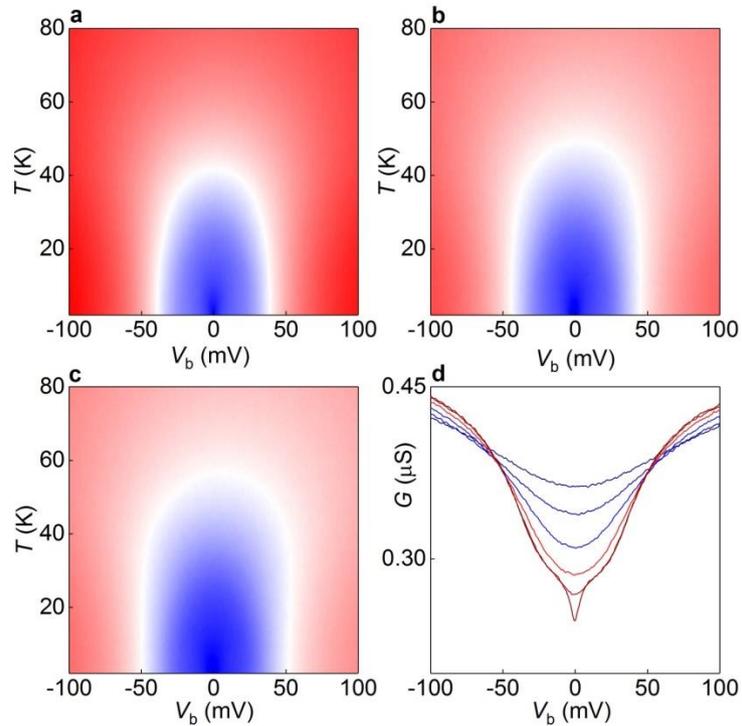

Fig. S2. **a**, Differential tunnel conductance as a function of $V_b$ and $T$ at $B_{||}$=0T and $V_g$=0V for a device with 2 layer thick CrBr$_3$ tunnel barrier (colour scale blue to white to red, 244μS to 343μS to 443μS for all contour plots in the figure). **b**, Similar to a, except $B_{||}$=10T. **c**, Similar to a, except $B_{||}$=17.5T. **d**, Zero-field tunnel conductance plots for various temperatures at a fixed gate voltage of $V_g$=0V (blue to purple curves range from $T$=2K to $T$=50K with the step 10K).

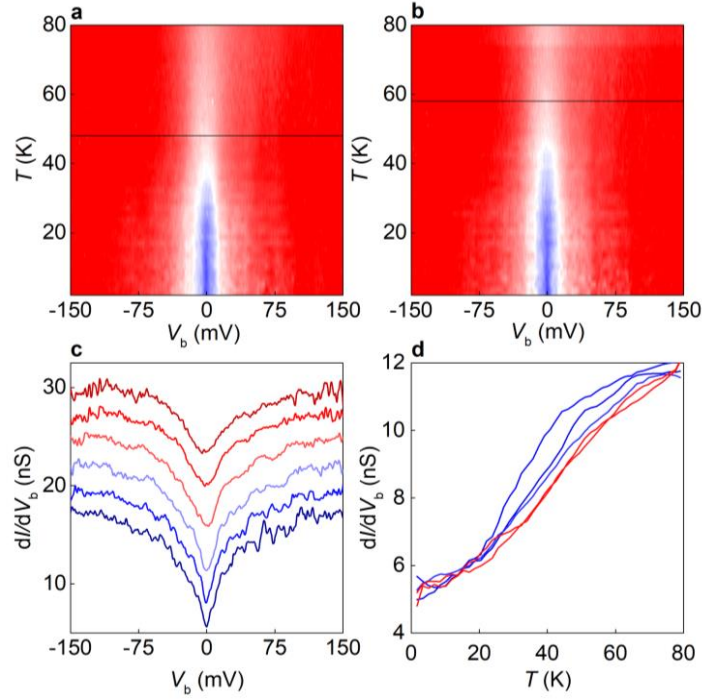

Fig. S3. **a**, Differential tunnel conductance as a function of $V_b$ and $T$ at $B_{||}$=0T and $V_g$=0V for a device with a 6-layer thick CrBr$_3$ tunnel barrier (colour scale blue to white to red, 3nS to 9.25nS to 16.5nS for all contour plots in the figure). Black horizontal lines in a,b mark the position of the extracted value of $T_c$. **b**, Similar to a, except $B_{||}$=10T. **c**, Zero-field tunnel conductance plots for various temperatures at a fixed gate voltage, $V_g$=0V (blue to purple curves range from 2K to 50K in steps of 10K, curves are shifted vertically for clarity). **d**, Dependence of the zero bias tunneling conductance d$I$/d$V_b$ on $T$ for different $B_{||}$ (red to blue curves are $B_{||}$=0T, 5T, 10T, 15T, 17.5T).

### S3. Quantum capacitance of *Gr/CrBr$_3$/Gr* devices

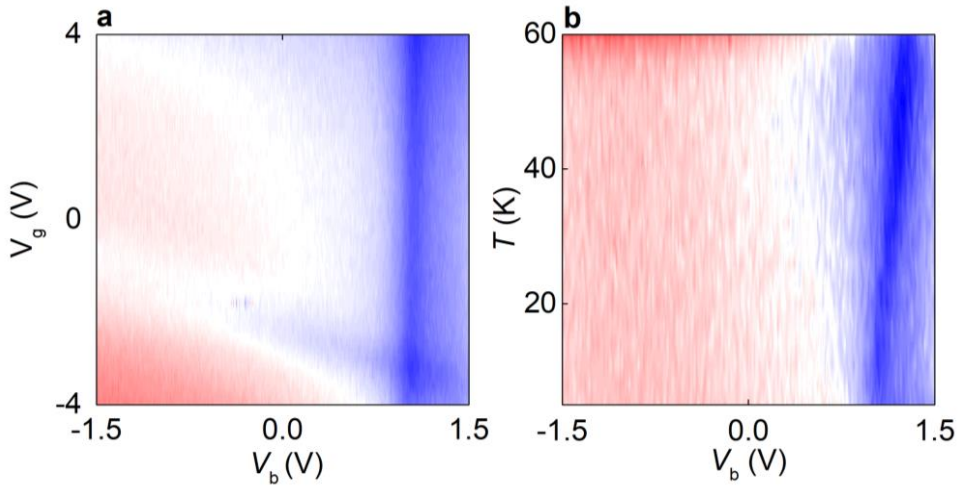

Fig. S4. Quantum capacitance measurements of the top gated *Gr/CrBr$_3$/Gr* device (fabricated on quartz substrate). CrBr$_3$ thickness 15nm, cross-section area 100μm$^2$. **a,** Capacitance as a function of $V_b$ and $V_g$ at $T$=2K and zero magnetic field (colour scale blue to white to red, 0.23pF to 0.24pF to 0.25pF). Two blue regions correspond to the suppressed density of states near the Dirac points of the top and bottom graphene layers. The edge of the blue region for the bottom layer is vertical, consistent with strong screening of the gate-induced electric field by the thick $CrBr_3$ barrier. **b,** the differential capacitance as a function of $V_b$ and $T$ at a fixed $V_g=0V$ voltage.

## S4. Calculation of magnon density of states

To calculate renormalized magnon spectra at finite temperatures we use self-consistent spin-wave theory in the form suggested in Ref.[S1]. We modify it to the case of a hexagonal lattice. For a single layer, there are two magnetic atoms per elementary cell with the corresponding magnon annihilation operators $b_1$ and $b_2$. The magnon spectrum is diagonalized by the same unitary transformation as for the electronic spectrum of graphene [S2],

$$b_{\vec{k}1} = \frac{1}{\sqrt{2}}(\alpha_{\vec{k}1} + \alpha_{\vec{k}2}),$$

$$b_{\vec{k}2} = \frac{S(\vec{k})}{\sqrt{2}f(\vec{k})}(\alpha_{\vec{k}1} - \alpha_{\vec{k}2})$$

where $\vec{k}$ is the magnon wave vector, $S(\vec{k}) = \sum_{\vec{\delta}} e^{i\vec{k}\vec{\delta}}$, $\vec{\delta}$ are the nearest neighbour vectors and $f(\vec{k}) = |S(\vec{k})|$. The self-consistent magnon spectrum at finite temperatures is determined by the two parameters, $\gamma, \gamma_L$ characterizing the renormalization of in-plane and out-of-plane exchange integrals ($J$ and $J_L$), respectively. The two magnon branches are:

$$E_{\vec{k}1,2} = 2J\gamma(3 \mp f(\vec{k})) + 2J_L\gamma_L(1 - \cos k_z c) - \mu,$$

where $c$ is the interlayer distance and $\mu$ is the magnon chemical potential which is nonzero above the Curie temperature. The self-consistent equations for the parameters $\gamma, \gamma_L$ and the average spin per site $\bar{S}$ are

$$\gamma = \bar{S} + \frac{1}{6}\sum_{\vec{k}} f(\vec{k})[N_B(E_{\vec{k}1}) - N_B(E_{\vec{k}2})]$$

$$\gamma_L = \bar{S} + \frac{1}{2}\sum_{\vec{k}}[N_B(E_{\vec{k}1}) + N_B(E_{\vec{k}2})]\cos k_z c$$

$$\bar{S} = S + (2S+1)N_B(E_f) - \frac{1}{2}\sum_{\vec{k}}[N_B(E_{\vec{k}1}) + N_B(E_{\vec{k}2})]$$

where $E_f = (2S+1)(6J\gamma + 2J_L\gamma_L - \mu)$ and $N_B(E) = \frac{1}{\exp(E/k_B T) - 1}$ is the Bose Einstein distribution function.

The computational results are shown in Fig.S5. The calculations show that the effective magnetic coupling between the layers characterized by the parameter $\gamma_L$ disappears near the Curie temperature whereas the short-range order in-plane characterized by the parameter $\gamma$ survives to higher temperatures, as expected for quasi-two-dimensional magnets [S3].

The in-plane correlation length $\xi(T)$ can be estimated from the effective spin-wave spectrum above the Curie temperature,

$$E(\vec{k}) = D\left(k^2 + \frac{1}{\xi^2}\right),$$

where $D$ is the spin-wave stiffness constant. For the particular case of the honeycomb lattice in the nearest-neighbour approximation

$$a/\xi(T) = \sqrt{\frac{2|\mu|}{3J\gamma}},$$

where $a$ is the distance between the neighbouring magnetic ions. The computational results are shown in Fig. S6.

The self-consistent spin-wave theory does not give a very accurate estimate of $T_c$, but is rather an upper estimate. Fluctuation corrections will lower the value of $T_c$ by a factor ~1.5 [S1]. Therefore, in our comparison of the theory with experiment, we have shifted the temperature dependence of calculated variables by the calculated value of $T_c$ but keeping the same ratio of $J/J_L$.

To consider how the magnetic field $B$ affects the magnon spectrum, one needs to replace $-\mu$ by the Zeeman gap $2\mu_B B$, where $\mu_B$ is the Bohr magneton, in all the above equations. Self-consistent solution of the equations for our parameters gives us the shift of magnon energy peaks with the slope being between 4.5·$\mu_B$ V/T and 7.1·$\mu_B$ V/T depends on two Van Hove singularities at temperature close to $T_c$ (Figure S7).

Within the self-consistent spin-wave theory, the intensity of the two-magnon processes (one magnon is emitted and one magnon is adsorbed) with the total energy change close to zero and the total wave vector change equal to $\vec{q}$ is proportional to the corresponding contribution to the longitudinal spectral density and reads [S4]

$$K_{\vec{q}}(T,H) = \sum_{\vec{k}} N_{\vec{k}}(1 + N_{\vec{k}+\vec{q}})\delta(E_{\vec{k}} - E_{\vec{k}+\vec{q}})$$

In all the calculations we use the experimental values of the exchange parameters. We have performed also the first-principle calculations of these quantities. Assuming the nearest-neighbour approximation, we estimated the parameters $J$ and $J_L$ from the total energy differences of various magnetic configurations (ferromagnetic, ferromagnetic in each layers with antiferromagnetic orientation between the layers, and antiferromagnetic within the layer) using the accurate full potential linearized augmented plane-wave (FLAPW) Wien2k code [S5] within the generalized gradient approximation (GGA) and experimental crystal structure. The results for (12x12x2) k-point mesh are $J$=0.94mV and $J_L$=0.076mV, respectively. The theoretical exchange $J_L$ parameter agrees well with the experimental values [S6], while the intralayer exchange $J$ is about 45% smaller. If one uses the theoretical GGA lattice parameters [S7] instead of the experimental ones, the corresponding exchange $J$ is about 44% larger than the experimental value [S6]. So strong dependence of the calculated exchange parameters on interatomic distances makes the use of the experimental values the most reliable choice.

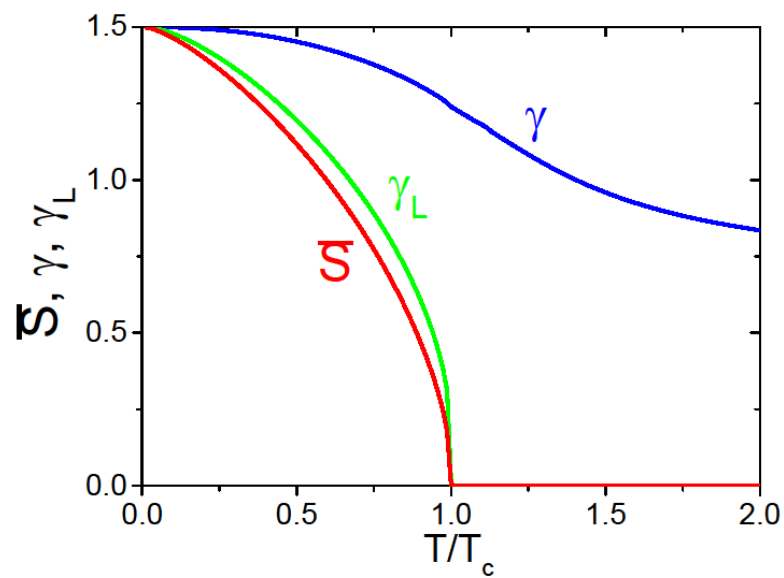

Figure S5. Temperature dependences of $\bar{S}, \gamma, \gamma_L$ at zero magnetic field.

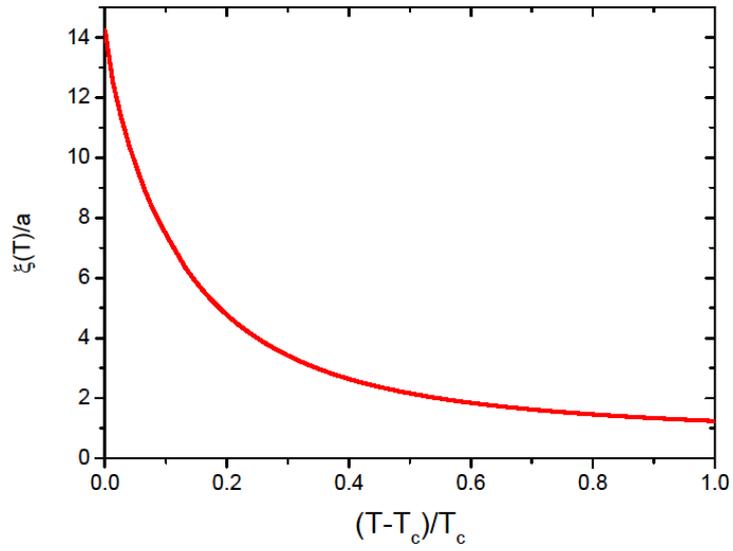

Figure S6. Temperature dependence of $\xi(T)/a$.

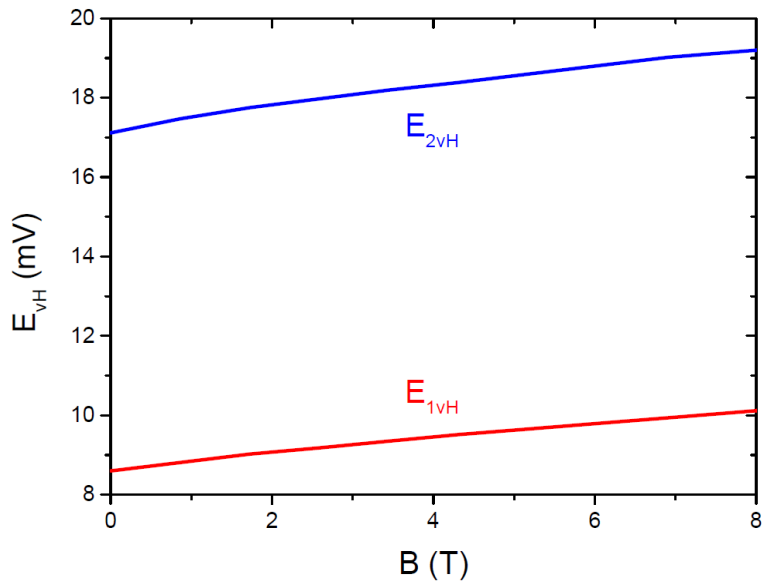

Figure S7. Magnetic field dependence of the positions of van Hove singularities in magnon spectrum.

**S5. Scattering rates**

The scattering rate for the two magnon (one magnon is emitted and one magnon is adsorbed) is proportional to the corresponding contribution to the longitudinal spectral density which is given by the function $K_q$. To make this relationship explicit we define $W_{2mag} \propto K_q$. The temperature dependence of $W_{2mag}$ is shown in Fig S8.

The scattering rate from the disordered spin texture at high temperatures $T \gtrsim T_c$ is proportional to the spin correlator for the longitudinal component of the atomic spin, $\langle \delta S^z \rangle^2$, see ref. 36 of the manuscript for more detail, so that

$$W_{spin} \propto \langle \delta S^z \rangle^2$$

We find that a good fit to the data can be obtained if we take a mean field approach where

$$\langle \delta S^z \rangle^2 = S(S+1) + \langle S^z \rangle \coth \beta/2 - \langle S^z \rangle$$

and the expectation value of the spin $\langle S^z \rangle$ is obtained using the standard self-consistent solution of the Brillouin function, using the Curie Weiss law so that:

$$\beta = \frac{gS\mu_B}{k_B T}(B + \lambda M)$$

In which $\mu_B$ is the Bohr magneton, $(Tc)$ is the Weiss molecular field constant and $M = Ng\mu S^z$ is the magnetisation.

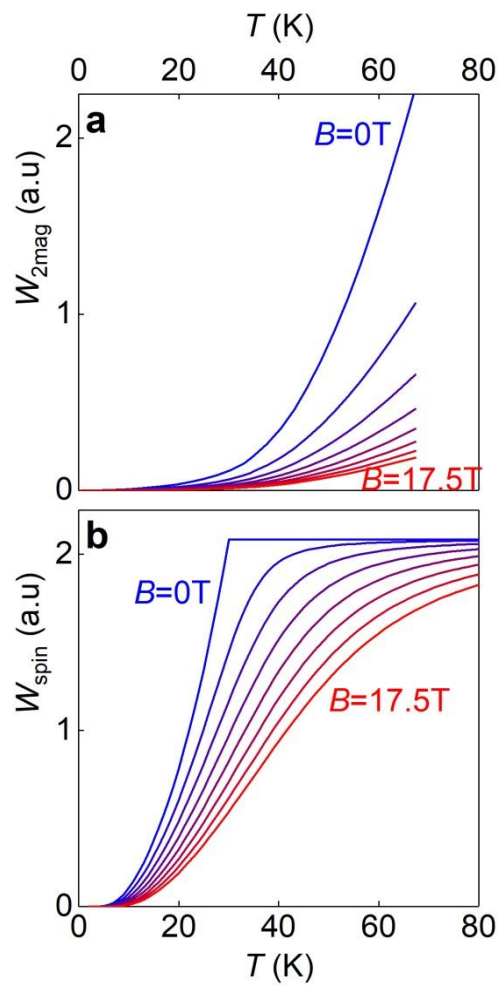

Figure S8. Scattering rates. Curves blue to red are for magnetic fields $B_\parallel=0T$ to $B_\parallel=17.5T$ in 2.5T increment.

**Supplementary References**